\documentclass[twocolumn,onecolappendix]{aastex63}
\usepackage{amsmath}

\shortauthors{Sheehan et al.}
\shorttitle{eDisk@VLA: L1527 IRS}

\begin{document}

\title{\bf A VLA View of the Flared, Asymmetric Disk Around the Class 0 Protostar L1527 IRS}

\author[0000-0002-9209-8708]{Patrick D. Sheehan}
\altaffiliation{NSF Astronomy \& Astrophysics Postdoctoral Fellow}
\affiliation{Center for Interdisciplinary Exploration and Research in Astronomy, 1800 Sherman Rd., Evanston, IL 60202, USA}

\author[0000-0002-6195-0152]{John J. Tobin}
\affiliation{National Radio Astronomy Observatory, 520 Edgemont Rd., Charlottesville, VA 22903, USA}

\author[0000-0002-7402-6487]{Zhi-Yun Li}
\affil{Astronomy Department, University of Virginia, Charlottesville, VA 22904, USA}

\author[0000-0002-2555-9869]{Merel L. R. van 't Hoff}
\affil{Department of Astronomy, University of Michigan, 1085 S. University Ave., Ann Arbor, MI 48109-1107, USA}

\author[0000-0001-9133-8047]{Jes K. J{\o}rgensen}
\affil{Niels Bohr Institute, University of Copenhagen, {\O}ster Voldgade 5-7, DK-1350 Copenhagen K., Denmark}

\author[0000-0003-4022-4132]{Woojin Kwon}
\affiliation{Department of Earth Science Education, Seoul National University, 1 Gwanak-ro, Gwanak-gu, Seoul 08826, Republic of Korea}
\affiliation{SNU Astronomy Research Center, Seoul National University, 1 Gwanak-ro, Gwanak-gu, Seoul 08826, Republic of Korea}

\author[0000-0002-4540-6587]{Leslie W. Looney}
\affiliation{Department of Astronomy, University of Illinois, 1002 West Green Street, Urbana, IL 61801, USA}

\author[0000-0003-0998-5064]{Nagayoshi Ohashi}
\affiliation{Academia Sinica Institute of Astronomy \& Astrophysics, 11F of Astro-Math Bldg, No.1, Sec. 4, Roosevelt Rd, Taipei 10617, Taiwan, R.O.C.}

\author[0000-0003-0845-128X]{Shigehisa Takakuwa}
\affiliation{Academia Sinica Institute of Astronomy \& Astrophysics, 11F of Astro-Math Bldg, No.1, Sec. 4, Roosevelt Rd, Taipei 10617, Taiwan, R.O.C.}
\affiliation{Department of Physics and Astronomy, Graduate School of Science and Engineering, Kagoshima University, 1-21-35 Korimoto, Kagoshima, Kagoshima 890-0065,
Japan}

\author[0000-0001-5058-695X]{Jonathan P. Williams}
\affil{Institute for Astronomy, University of Hawaii, 2680 Woodlawn Drive, Honolulu, HI 96822, USA}

\author[0000-0002-8238-7709]{Yusuke Aso}
\affil{Korea Astronomy and Space Science Institute (KASI), 776 Daedeokdae-ro, Yuseong-gu, Daejeon 34055, Republic of Korea}

\author[0000-0001-5782-915X]{Sacha Gavino}
\affil{Niels Bohr Institute, University of Copenhagen, {\O}ster Voldgade 5-7, DK-1350 Copenhagen K., Denmark}

\author[0000-0003-4518-407X]{Itziar de Gregorio-Monsalvo}
\affil{European Southern Observatory (ESO), Alonso de C\'{o}rdova 3107, Vitacura, Santiago 763-0355, Chile}

\author{Ilseung Han}
\affil{Korea Astronomy and Space Science Institute (KASI), 776 Daedeokdae-ro, Yuseong-gu, Daejeon 34055, Republic of Korea}
\affil{University of Science and Technology, Korea (UST), 217 Gajeong-ro, Yuseong-gu, Daejeon 34113, Republic of Korea}

\author[0000-0002-3179-6334]{Chang Won Lee}
\affil{Korea Astronomy and Space Science Institute (KASI), 776 Daedeokdae-ro, Yuseong-gu, Daejeon 34055, Republic of Korea}
\affil{University of Science and Technology, Korea (UST), 217 Gajeong-ro, Yuseong-gu, Daejeon 34113, Republic of Korea}

\author[0000-0002-9912-5705]{Adele Plunkett}
\affiliation{National Radio Astronomy Observatory, 520 Edgemont Rd., Charlottesville, VA 22903, USA}

\author[0000-0002-0549-544X]{Rajeeb Sharma }
\affil{Niels Bohr Institute, University of Copenhagen, {\O}ster Voldgade 5-7, DK-1350 Copenhagen K., Denmark}

\author[0000-0003-3283-6884]{Yuri Aikawa}
\affil{Department of Astronomy, Graduate School of Science, The University of Tokyo, 7-3-1 Hongo, Bunkyo-ku, Tokyo 113-0033, Japan}

\author[0000-0001-5522-486X]{Shih-Ping Lai}
\affiliation{Institute of Astronomy and Department of Physics, National Tsing Hua University, No. 101, Section 2, Kuang-Fu Road, Hsinchu 30013, Taiwan}
\affiliation{Academia Sinica Institute of Astronomy \& Astrophysics, 11F of Astro-Math Bldg, No.1, Sec. 4, Roosevelt Rd, Taipei 10617, Taiwan, R.O.C.}

\author[0000-0003-3119-2087]{Jeong-Eun Lee}
\affil{School of Space Research, Kyung Hee University, 1732, Deogyeong-Daero, Giheung-gu Yongin-shi, Gyunggi-do 17104, Korea}

\author{Zhe-Yu Daniel Lin}
\affil{Astronomy Department, University of Virginia, Charlottesville, VA 22904, USA}

\author[0000-0003-1549-6435]{Kazuya Saigo}
\affil{National Astronomical Observatory of Japan (NAOJ), National Institutes of Natural Sciences (NINS), 2-21-1 Osawa, Mitaka, Tokyo 181-8588, Japan}

\author[0000-0001-8105-8113]{Kengo Tomida}
\affil{Astronomical Institute, Tohoku University, Aoba, Sendai, Miyagi 980-8578, Japan}

\author[0000-0003-1412-893X]{Hsi-Wei Yen}
\affiliation{Academia Sinica Institute of Astronomy \& Astrophysics, 11F of Astro-Math Bldg, No.1, Sec. 4, Roosevelt Rd, Taipei 10617, Taiwan, R.O.C.}

\begin{abstract}
We present high resolution Karl G. Jansky Very Large Array (VLA) observations of the protostar L1527 IRS at 7 mm, 1.3 cm, and 2 cm wavelengths. We detect the edge-on dust disk at all three wavelengths and find that it is asymmetric, with the southern side of the disk brighter than the northern side. We confirm this asymmetry through analytic modeling and also find that the disk is flared at 7 mm. We test the data against models including gap features in the intensity profile, and though we cannot rule such models out, they do not provide a statistically significant improvement in the quality of fit to the data. From these fits, we can however place constraints on allowed properties of any gaps that could be present in the true, underlying intensity profile. The physical nature of the asymmetry is difficult to associate with physical features due to the edge-on nature of the disk, but could be related to spiral arms or asymmetries seen in other imaging of more face-on disks.
\end{abstract}


\section{Introduction}

Protostellar disks are a natural consequence of conservation of angular momentum during the star formation crocess when the natal cloud collapses to form a young star \citep[e.g.][]{Terebey1984,Ulrich1976}. Such disks are thought to be established early in this process \citep[e.g.][]{Tobin2012,Eisner2012,Sheehan2017,Tobin2020,Garufi2021ALMAEmission} and are important for setting the stage for planet 
formation \citep[e.g.][]{Sheehan2017,Sheehan2018,Tychoniec2020,Segura-Cox2020FourOld}. 
Recent high resolution imaging of protoplanetary disks, the more evolved siblings of protostellar disks, has uncovered a diversity of ``substructures", features that deviate from an otherwise smooth, monotonically decreasing intensity profile. These substructures typically come in the form of narrow rings and gaps \citep[e.g.][]{ALMAPartnership2015FirstRegion,Isella2016,Andrews2016,Long2018,Huang2018,Cieza2021TheResolution,Sierra2021MoleculesEmission} that are typically associated with the depletion of dusty material in certain regions of disks, but also as large scale asymmetries \citep[e.g.][]{VanDerMarel2013,Casassus2013,Boehler2018TheCavity,Cazzoletti2018EvidenceDisk,Dong2018TheDisk} and spiral arms \citep[e.g.][]{Perez2016,Huang2018a,Huang2021MoleculesDisk}. 

Disk substructures are frequently tied to the presence of planets hiding in the disks, and carving the material \citep[e.g.][]{Dodson-Robinson2011,Kley2011Planet-DiskEvolution,Zhu2012,Dong2015,Zhang2018TheInterpretation}, leading many to use these features as signposts of young planets. Few planets have, however, as of yet been found hiding within substructures directly \citep[e.g.][]{Keppler2018}, though the presence of some have been inferred through kinematic features \citep[e.g.][]{Teague2018A163296,Pinte2018KinematicDisk,Isella2019DetectionProtoplanets,Pinte2020NineGaps}. As such, a number of other mechanisms have therefore been proposed to explain their presence \citep[e.g.][]{Cuzzi2004,Zhang2015,Flock2015,Okuzumi2016,Suriano2018,Takahashi2018,Ohashi2021RingProtostar}. Regardless of whether they are carved by planets or other mechanisms, substructures likely represent current over-densities of dust that may be conducive to further planet formation within.

The ubiquity of substructures in more-evolved disks \citep[e.g.][]{Huang2018,Andrews2018,Long2018} begs the question of when substructures first arise in disks. As a result of the connection between substructures and planets or planet formation, a few corollaries to this question are when planet formation begins, and at what time might planets or planetary embryos be hiding in disks. Some evidence exists that such features may be present in Class I protostellar disks \citep{Sheehan2017a,Sheehan2018a,Sheehan2020, Segura-Cox2020FourOld,DeValon2020}, but whether substructure can be present in the earliest phase of star-formation, the Class 0 phase \citep[e.g.][]{Andre1993} is as of yet uncertain. Though spiral arms have been identified in such young disks \citep[e.g.][]{Tobin2016,Lee2020,Takakuwa2020CircumbinaryRegion}, those sources are multiple systems, and the features are likely the result of companion formation in gravitationally unstable disks, gravitational interactions with a circum-multiple disk, or accretion from the envelope onto the disk, rather than the direct impact of ongoing planet formation.

In this paper, we present new, high resolution and high sensitivity observations of L1527 IRS
taken with the National Science Foundation's Karl G. Jansky Very Large Array (VLA) at 7 mm, 1.3 cm, and 2 cm, as a part of the Early Planet Formation in Embedded Disks with the VLA (eDisk@VLA) program, a companion to the forthcoming Atacama Large Millimeter Array (ALMA) Large Program of the same name (Ohashi et al., in prep). The goal of the eDisk Large Program is to conduct a systematic search for substructures in young \citep[$\lesssim1$ Myr-old; e.g.][]{Evans2009,Dunham2015} disks to search for evidence of the early onset of planet formation. The companion VLA program is meant to image many of those same sources at long wavelengths in order to help constrain dust grain properties and better characterize substructures at wavelengths where optical depth should be less significant. 

L1527 IRS is a well-known edge-on \citep[e.g.][]{Ohashi1997,Tobin2008ConstrainingModeling}, single \citep[][c.f. \citet{Loinard2002Orbital04368+2557}]{Nakatani2020} Class 0 protostar in the Taurus star-forming region and has been extensively studied across a range of wavelengths. Though formally classified as a Class 0 protostar \citep[e.g.][]{Kristensen2012WaterProtostars}, it is difficult to tie this classification to an evolutionary stage for the system due to the edge-on geometry of the disk \citep[e.g.][]{Crapsi2008}. Though ages of protostars are notoriously difficult to measure, a more physically motivated system for classifying the evolutionary stage of a system, such as considering how the envelope mass compares with the protostellar mass \citep[e.g.][]{Robitaille2006}, suggests that L1527 IRS is indeed quite young, even if it is not among the youngest of the Class 0 systems \citep[see e.g.][and references therein]{Tobin2013}.

Early observations with the VLA suggested that the disk is likely optically thick out to at least 1 mm, but that there could be significant optically thick material even at wavelengths of $\sim1$ cm \citep{Melis2011MICROWAVERESULTS}. It was the first Class 0 protostellar source identified to have a Keplerian-rotation-supported disk \citep{Tobin2012,ohashi2014}. \citet{Aso2017} estimated a protostellar mass of 0.45 M$_{\odot}$ and a disk radius of 74 au from further ALMA kinematic observations. The disk is also warm, with midplane temperatures exceeding 20 K out to $\gtrsim75$ au \citep{VanTHoff2018} based on the presence of CO emission at large radii, and molecular line observations also show interesting features near the disk-envelope transition \citep[e.g.][]{Sakai2014a,Sakai2014}. Finally, and most relevant to the work presented here, it was recently suggested to have substructures in the form of three clumps spread across the disk \citep{Nakatani2020} from earlier observations with the VLA at 7~mm.

The structure of this work is as follows: we describe the observations and data reduction in Section \ref{section:observations}. In Section \ref{section:analysis}, we present our observations and perform careful modeling using analytic intensity profiles to characterize the structure of the disk at each wavelength included in our observations. Finally, we discuss the implications of the disk structure found as a result of our modeling in Section \ref{section:discussion}, and summarize our results in Section \ref{section:conclusion}. We further apply the same modeling framework to the observations of L1527 IRS at 7 mm reported in \citet{Nakatani2020} to test how our results compare with that work in Appendix \ref{section:nakatani_comparison}.

\section{Observations \& Data Reduction}
\label{section:observations}

L1527 IRS was observed by the VLA in four epochs between 7 January 2021 and 14 January 2021 (Program 20B-322) in the A-configuration, with baselines ranging from 680 m -- 36.4 km. The pointing center for the observations was set at $\alpha$(J2000)$\,=\,04^{\rm h}39^{\rm m}53.9^{\rm s}$ $\delta$(J2000)$\,=\,26^{\circ}03{\arcmin}09.6{\arcsec}$ based on \citet{VanTHoff2018}. The observations were taken using the Q-band (44 GHz, 7 mm), K-band (22 GHz, 1.3 cm), and Ku-band (15 GHz, 2 cm) receivers. The Q-band and K-band observations were taken during the same epochs using band switching and were observed three times on 7, 8, and 12 January 2021, while the Ku-band observations were taken during a single epoch on 14 January 2021. The Q-band and K-band receivers were configured in wideband continuum mode, with four 2 GHz wide basebands centered at 41, 43, 45, 47 GHz and 19, 21, 23, 25 GHz, respectively. The Ku-band receivers were also configured in wideband continuum mode but with only three 2 GHz wide basebands centered at 13, 15, 17 GHz.

\begin{figure*}
    \centering
    \includegraphics[width=7in]{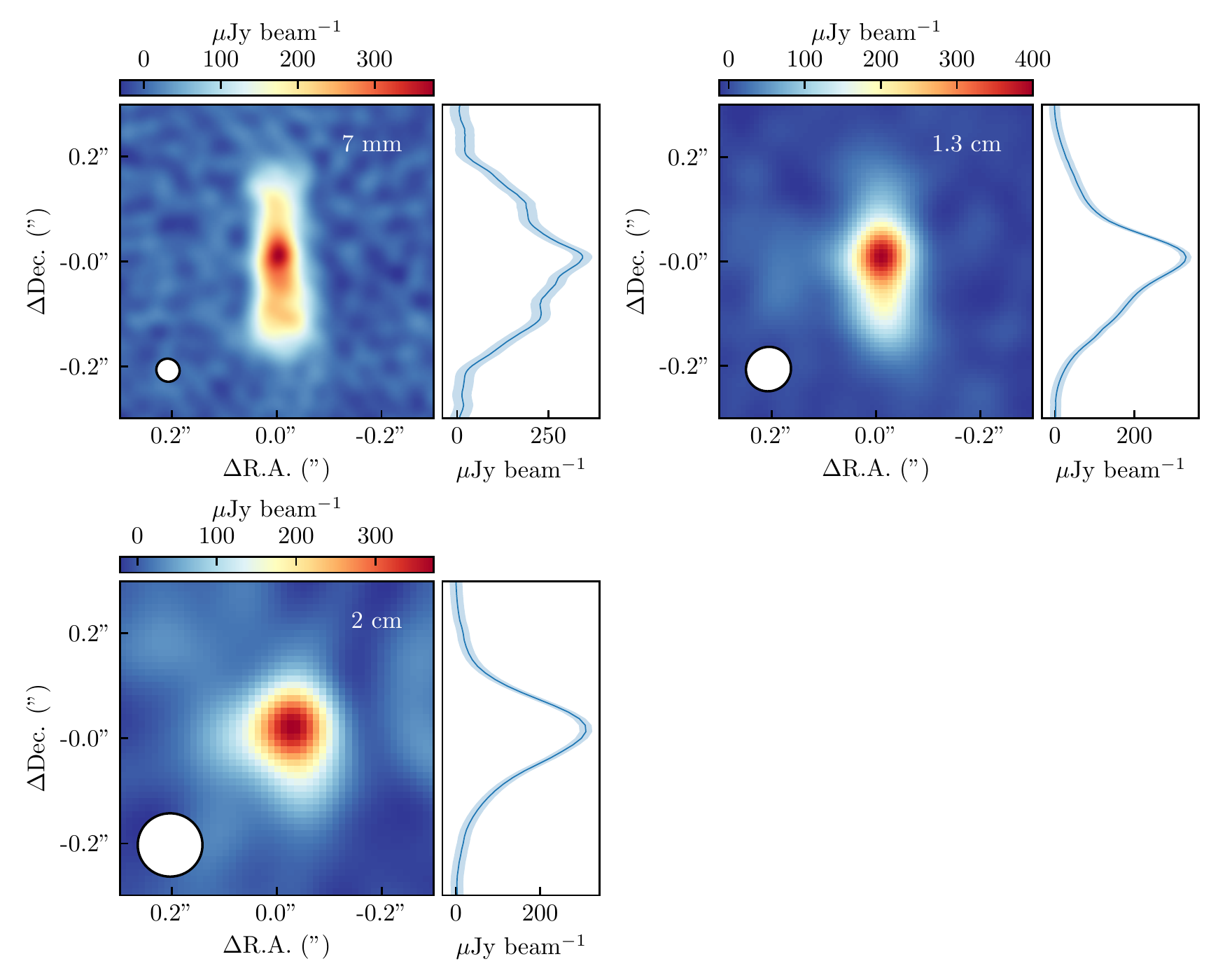}
    \caption{Robust=0.5 weighted images of our L1527 IRS VLA observations at 7 mm ({\it top left}), 1.3 cm ({\it top right}), and 2 cm ({\it bottom left}). To the right of each image we show the intensity profile from a one-dimensional pixel slice in the North-South direction through the center of the disk, along with a shaded region representing the $3\sigma$ confidence interval. The edge-on disk is clearly visible at 7 mm but also can be made out at longer wavelengths, as well. An asymmetry in the disk, with the southern side brighter than the northern side, can also be made out in the all three images, though is most prominent at 7 mm and 1.3 cm.}
    \label{fig:data}
\end{figure*}

The science target was observed along with the quasars 3c147 as the flux calibrator and 3c84 as the bandpass calibrator for all bands. For complex gain calibration, the quasar J0440+2728 was used as the calibrator for Q and K-band, and J0403+2600 was the phase calibrator in Ku-band. The total time on-source for L1527 IRS was 190 minutes in Q-band, 50 minutes in K-band, and 18 minutes in Ku-band. The data were reduced, including flux, bandpass and phase calibration along with automatic flagging for radio-frequency interference (RFI) using the VLA pipeline in the CASA software package \citep{Mcmullin2007}, version 6.1.2. If additional flagging was found
to be necessary after a pipeline run, the necessary flags were applied to the data and the
pipeline was re-run.

For all bands, we image the data using the {\it tclean} routine within CASA using multi-frequency synthesis mode and a robust parameter of 0.5. We did repeat the imaging using a range of robust parameters and found that 0.5 produced the best balance between resolution and sensitivity, but also note that the bulk of the remainder of our analysis is done in the $uv$-plane and is therefore unaffected by this choice. The resulting Q-band continuum image has a beam size of 0\farcs045 $\times$ 0\farcs043 with a position angle of 51.6$^{\circ}$ and an RMS of 8.8 $\mu$Jy beam$^{-1}$. The K-band continuum image has a beam size of 0\farcs087 $\times$ 0\farcs084 with a position angle of -58.0$^{\circ}$ and an RMS of 4.6 $\mu$Jy beam$^{-1}$. Finally, the Ku-band continuum image has a beam size of 0\farcs123 $\times$ 0\farcs120 with a position angle of -89.5$^{\circ}$ and an RMS of 5 $\mu$Jy beam$^{-1}$. We show these images in Figure \ref{fig:data}. We note that there is a small systematic spatial offset of $\sim0\farcs035$ between our Q and K-band data with respect to the Ku-band data. This is due to using a different phase calibrator at
Ku-band data relative to Q and K-bands. This offset does not, however, affect our analysis as we consider each wavelength separately.

As much of our analysis will be done in the $uv$-plane, we check the uncertainties on the visibilities by comparing $\sigma_{vis} = 1 / \sqrt{\Sigma ( 1 / \sigma_i^2 )}$, where $1/\sigma_i^2 = w_i$ is the weight for an individual integration as given by the VLA pipeline, with the RMS of a naturally weighted image, and scale the weights on the visibilities until they match. Though it is not necessarily the case that $\sigma_{vis}$ must match the RMS of a naturally weighted image exactly, as systematic uncertainties and imaging artefacts could affect this comparison, they should be in reasonable agreement. This results in scaling the weights on the visibilities, $w_i$, by a factor of 0.125. We note that the correction factor is small, which also increases the uncertainties, and is, therefore, a conservative estimate of the uncertainties on the data. 

Finally, there is an additional uncertainty on the overall flux calibration of the data, typically on the order of 10\% of the flux. We do not, however, consider this uncertainty in our analysis as it only affects the flux scale of separate observations relative to each other, and does not have an impact on the uncertainties within an observation, i.e. on source structure. Any flux density uncertainties reported within this work, however, are purely statistical, so one should include an additional 10\% uncertainty on reported fluxes when used beyond this work.

\section{Analysis \& Results}
\label{section:analysis}

In agreement with previous observations \citep{Ohashi1997,Loinard2002Orbital04368+2557,Tobin2008ConstrainingModeling,Tobin2010TheImaging,Tobin2012,Tobin2013,Sakai2014,Aso2017,Nakatani2020}, our new data (shown in Figure \ref{fig:data}) of L1527 IRS show an edge-on disk elongated in the North-South direction. This can be seen clearly at 7 mm, but the North-South disk can also be seen peeking out from the bright central point source at 1.3 and 2 cm. The extent of the disk at 7 mm is consistent with previous 7 mm observations by \citet{Loinard2002Orbital04368+2557} and \citet{Nakatani2020}, while it is significantly smaller than the disk extent at shorter millimeter wavelengths \citep[e.g.][]{Sakai2014,Aso2017,Nakatani2020}. This discrepancy is likely due to the radial drift \citep[e.g.][]{Weidenschilling1977AerodynamicsNebula} and/or preferential growth \citep[e.g.][]{Birnstiel2010Gas-andDisks} of the larger dust grains traced by these long wavelength observations.

The bright central emission seen at all three wavelengths is likely a combination of dust and free-free emission from the protostellar jet, with the fraction increasing with wavelength. Indeed, a weak East-West protrusion can be made out in both the 1.3 and 2 cm observations. Moreover, the spectral index of the central peak in each of our images, as calculated from a simple two-dimensional Gaussian fit limited to the central region of the emission at each wavelength, is consistently $<1.5$ and is equal to $\sim0.8$ between our longest wavelength images, indeed suggesting some amount of free-free emission mixed with the dust thermal emission.

We find that in both the 7 mm and 1.3 cm images, the disk appears to be asymmetric with the southern half of the disk brighter than the northern half. At 7mm the difference in intensity is approximately $3\sigma$ when measured $0\farcs1$ north (204 $\mu$Jy beam$^{-1}$) and south (232 $\mu$Jy beam$^{-1}$) of the emission peak, though the emission is resolved so this may be more significant when integrated over larger regions. At 1.3 cm the asymmetry is more pronounced, with an intensity of 190 $\mu$Jy beam$^{-1}$ measured $0\farcs1$ south of the peak and 98 $\mu$Jy beam$^{-1}$ to the north, a difference of $\sim20\sigma$. This asymmetry may also be present in the 2 cm observations, particularly visible on the southern side of the disk, but is more difficult to pick out by-eye. In contrast with \citet{Nakatani2020}, we find that the disk has no apparent large, clumpy structures; as our new observations have $\sim10\times$ higher sensitivity than the images previously presented, such features should have been present at the $>10\sigma$ level. Instead, we find that the northern and southern clumps previously identified correspond to the northern and southern shoulders of the disk in our imaging. For interested readers, we compare our results with their work in greater detail in Appendix \ref{section:nakatani_comparison}. On the southern half of the disk at 7 mm, around $\sim0.1''$ south of the central peak, there is a weak feature that might deviate from the smoothly decreasing emission. From the images, however, it is not immediately obvious whether this feature is real, and if so, what it might represent. If we consider the difference between the lowest and highest values found near the feature in the image plane, then the significance is $<3\sigma$, and so quite tentative.

To better characterize the disk emission at all three wavelengths and determine whether such features seen in these images are real in a statistically rigorous way, we fit analytic models to our observational data. We describe the model, fitting procedure, and results below.

\subsection{Analytic Models}
\label{section:modeling}

To characterize the brightness distribution of L1527 IRS in each image, we fit analytic models to the visibility data to try and reconstruct the image. The goal of this modeling is not to necessarily provide a fully physically motivated model, such as a radiative transfer model \citep[e.g.][]{Sheehan2017}, but rather to produce a simple model that can reproduce the features seen in the image such that we can test whether those features are statistically significant in a more quantitative way. We provide further interpretation of these model features in Section \ref{section:discussion}. We start with a simple model but add additional components motivated by the features seen in each image.

To find the best-fit set of parameters for each model, we use the \texttt{dynesty} package \citep{Speagle2019} to sample the posterior distributions using Nested Sampling \citep{Skilling2004NestedSampling,Skilling2006NestedComputation}. Nested Sampling is a method for estimating the Bayesian Evidence for a model. Bayesian Evidence, or the marginal or integrated likelihood, is given by
\begin{equation}
    Z_M = \int_{\Omega_{\Theta}} P(D|\Theta,M) P(\Theta|M) d\Theta,
\end{equation}
where $P(D|\Theta,M)$ is the likelihood of the data given the parameters $\Theta$ in the model $M$, $P(\Theta|M)$ is the prior for the parameters in the model, and $\Omega_{\Theta}$ represents the entire parameter space \citep[for further details see][]{Speagle2019}. The Bayesian Evidence is an important tool in model selection, as it provides a quantitative way to compare models through the Bayes Factor, or the ratio of Bayesian Evidences, i.e. Bayes Factor $= Z_{M_1}/Z_{M_2}$. Nested sampling calculates the Bayesian Evidence by sampling randomly from increasingly small nested shells of constant likelihood, and integrating the prior over these nested shells.\footnote{For further details, a nice description can be found in \citet{Speagle2019}.} By using Nested Sampling, we can simultaneously sample the posterior for our models, but also calculate the Bayesian Evidence to allow us to quantitatively compare the quality of fit of our different model choices.

Our base model is a rectangle, motivated by an inspection of the data in Figure \ref{fig:data}, in which the 7 mm data appears to be broadly rectangular in outline. A simple rectangular model would be that of a two-dimensional top hat, but it can have significant ringing due to the sharp edges, so we introduce an exponential taper in both directions to smooth the profile and minimize such effects. The analytic prescription for this rectangle is therefore given by,
\begin{equation}
    \label{equation:rectangle}
    I_{r} = I_0 \, \exp\left\{ -\frac{(x - x_0)^{\gamma_x}}{2 \, x_w^{\gamma_x}} - \frac{(y - y_0)^{\gamma_y}}{2 \, y_w^{\gamma_y}}\right\},
\end{equation}
where  the $x$ coordinate is defined to be along the major axis of the rectangle, and the $y$ coordinate is defined to be along the minor axis of the rectangle. In our base model, we fix $\gamma_x = \gamma_y = 4$ as this prescription provides a sharper truncation and a more rectangular appearance than a more traditional two-dimensional Gaussian function ($\gamma_x = \gamma_y = 2$), which would appear more like an oval. We do, however, allow both to vary, independently, in subsequent fits to allow for the possibility of smoother or sharper truncation. We also vary the centroid of the rectangle ($x_0$, $y_0$), and the width of the rectangle along the major ($x_w$) and minor ($y_w$) axes. Finally, though the emission is close to due North-South, we also allow the position angle ($p.a.$) of the rectangle to shift to match the slight offset.

To reproduce the North-South emission profile, which is sharply peaked at the center and not uniform across the extent of the disk, we add a power-law brightness profile along the major axis. We initially use a single power-law,
\begin{equation}
    I_{pl} = I_{r} \left(\frac{x'}{x_w}\right)^{-\gamma},
\end{equation}
where $\gamma$ is the power-law slope and is allowed to vary. To smooth between a point-source-like central component, where $I_{\nu} \longrightarrow \infty$ as $x \longrightarrow 0$, or a broader central peak, we add a small constant value to $x$, such that $x' = x + \delta$. We initially fix $\delta = 0.1 \, dx$, where $dx$ is the pixel size in the model image, but also allow it to vary in later fits. After an initial round of fitting, we also explored using a smoothly broken power-law profile,
\begin{equation}
    \label{equation:broken_powerlaw}
    I_{bpl} = I_{r} \left(\frac{x'}{x_w}\right)^{-\gamma_{in}} \, \left\{\frac{1}{2}\left[1 + \left(\frac{x'}{x_b}\right)^{1/\Delta}\right]\right\}^{(\gamma_{in} - \gamma_{out})\Delta},
\end{equation}
to better fit the central peak. Here, $\gamma_{in}$ and $\gamma_{out}$ are the inner and outer power-law slopes and $x_b$ is the point where the transition from inner to outer slope occurs. $\Delta$ controls the transition smoothness, with small values indicating a sharp transition and large values a slow, smooth transition. 

To test whether the disk is asymmetric, we allow the model to have differing power-law indices for $x > 0$ (south, towards the excess emission) and $x < 0$ (north). For the power-law model, this means that instead of $\gamma$, we have $\gamma_+$ and $\gamma_-$ parameters. For the broken-power law model, because the central peak appears to be more or less symmetric, we fix $\gamma_{in,+} = \gamma_{in,-}$, but allow the outer power-law indices to vary independently as $\gamma_{out,+}$ and $\gamma_{out,-}$.

We also note that the disk appears to be flared in the 7 mm image. To model this flaring, we allow the width of the rectangle in the y-direction to vary as a function of position in the x-direction,
\begin{equation}
    y_w = y_{w,0} \, \left[1 + A \left(\frac{x'}{x_w}\right)\right].
\end{equation}
Here, $A$ controls how much wider the disk is at $x = x_w$ compared with at $x = 0$, with the width scaling linearly along the major axis.

Though our observations presented in Figure \ref{fig:data} ostensibly show a disk with no clear gaps in the intensity profile, with the exception of the tentative feature on the southern side of the disk, visibility data can encode information at smaller spatial scales than are recovered by CLEANed images \citep[e.g.][]{Jennings2022AAu}. Moreover, clumps that appear gap-like were previously reported in 7 mm imaging of L1527 by \citet{Nakatani2020}. Therefore, to search for substructures, we add a gap to the model. To prevent ringing in the model from sharply truncating the intensity in the gap, we use a smooth gap model parameterized as a Gaussian subtracted from the gap-free model.
\begin{equation}
    I_{gapped} = I_{bpl} \, \left\{1 - (1 - \Delta_{gap}) \, \exp\left[-\frac{(x' - x_{gap})^2}{2 \, {w_{gap}}^2}\right]\right\}.
\end{equation}
Here $x_{gap}$ represents the center of the gap, $\Delta_{gap}$ the multiplicative factor by which the intensity is reduced at the center of the gap, i.e.,
\begin{equation}
I_{gapped}\Bigg\rvert_{x' = x_{gap}} = I_{bpl}\Bigg\rvert_{x' = x_{gap}} \Delta_{gap},
\end{equation}
and $w_{gap}$ the width of the gap. We did also consider other prescriptions for the gap, such as a simple one in which the density is reduced by $\Delta_{gap}$ within $|x' - x_{gap}| < w_{gap} / 2$, and found consistent results regardless of our exact choice of how to represent such a feature. We consider models with just a single gap on one half of the disk, motivated by the feature seen in the southern half of the disk in the 7 mm image, but also a model in which the gap feature is symmetric across the center of the disk.

\begin{deluxetable*}{cclc}
\tablenum{1}
\tabletypesize{\scriptsize}
\tablecaption{Summary of Analytic Model Parameters and Priors}
\label{table:analytic_priors}
\tablehead{\colhead{Parameter} & \colhead{Unit} & \colhead{Description} & \colhead{Prior}}
\startdata
$x_0$ & \arcsec & Center of the model in the East-West direction, with positive $x_0$ to the East & $x_{0,guess}-0.3 < x_0 < x_{0,guess} + 0.3$ \\ [2pt]
$y_0$ & \arcsec & Center of the model in the North-South direction, with positive $y_0$ to the North & $y_{0,guess}-0.3 < y_0 < y_{0,guess} + 0.3$ \\ [2pt]
$x_w$ & \arcsec & Width of the rectangle along the major axis & $\log_{10}{0\farcs005} < \log_{10}{x_w} < \log_{10}{1\arcsec}$ \\ [2pt]
$y_w$ & \arcsec & Width of the rectangle along the minor axis & $\log_{10}{0\farcs005} < \log_{10}{y_w} < \log_{10}{x_w}$ \\ [2pt]
$\gamma_{x}$ & \nodata & How smoothly/sharply the rectangle is tapered beyond $x_w$ & $2 < \gamma_{x} < 6$ \\ [2pt]
$\gamma_{y}$ & \nodata & How smoothly/sharply the rectangle is tapered beyond $y_w$ & $2 < \gamma_{y} < 6$ \\ [2pt]
$A$ & \nodata & How flared the disk is along the minor axis & $0 < A < 5$ \\ [2pt]
$\gamma_{in}$ & \nodata & Intensity power-law index for the inner region & $0 < \gamma_{in} < 2$ \\ [2pt]
$\gamma_{out,+}$ & \nodata & Intensity power-law index for the outer region in the positive-x direction & $-1 < \gamma_{out,+} < \gamma_{in}$ \\ [2pt]
$\gamma_{out,-}$ & \nodata & Intensity power-law index for the outer region in the negative-x direction & $-1 < \gamma_{out,-} < \gamma_{in}$ \\ [2pt]
$x_b$ & \arcsec & Position where the break from $\gamma_{in}$ to $\gamma_{out}$ occurs & $\log_{10}{0\farcs005} < \log_{10}{x_b} < \log_{10}{x_w}$ \\ [2pt]
$\log_{10}{\Delta}$ & \nodata & Length scale for transition from $\gamma_{in}$ to $\gamma_{out}$ & $-2 < \log_{10}{\Delta} < 2$ \\ [2pt]
$x_{gap}$ & \arcsec & Location of the gap along the major axis & $x_b < x_{gap} < x_w$ \\ [2pt]
$w_{gap}$ & \arcsec & Width of the gap along the major axis & $\log_{10}{0\farcs001} < \log_{10}{w_{gap}} < -1$ \\ [2pt]
$\Delta_{gap}$ & \nodata & Multiplicative reduction of the intensity within the gap & $-3 < \log_{10}{\Delta_{gap}} < 0$ \\ [2pt]
$p.a.$ & $^{\circ}$ & Position angle of the major axis of the rectangle, East of North. When the disk & $135^{\circ} < r_{c,r} < 225^{\circ}$ \\ 
  &  & is asymmetric, $p.a.$ references the direction of the brighter side. & \\ [2pt]
$F_{\nu}$ & Jy & Integrated flux of the disk-component of the model, in Jy & $-4 < \log_{10} F_{\nu} < -1$ \\ [2pt]
$\sigma_{env}$ & \arcsec & 1$\sigma$ radius of the large scale Gaussian envelope & $\log_{10}{x_w} < \log_{10}{\sigma_{env}} < \log_{10}{100\arcsec}$ \\ [2pt]
$F_{\nu,env}$ & Jy & Integrated flux of the large scale Gaussian & $-3 < \log_{10} F_{\nu,env} < 1$ \\ [2pt]
\enddata
\end{deluxetable*}

Rather than fit for the peak intensity, or intensity normalized to a specific location in the disk, we instead integrate the emission from the entire model over all space and re-scale the model image to a given total flux, $F_{\nu}$. We then, in our model fitting, use $F_{\nu}$ as a free parameter to report the total flux in each image.

Finally, in our initial fits we found that the shortest baseline data could typically not be fit well with the models that reproduce the disk emission alone. This is likely due to the presence of a low surface-brightness envelope around a young protostar that cannot be otherwise seen below the noise in the image. To account for this emission in our model, we also add a simple, large scale envelope component represented by a symmetric, two dimensional Gaussian with total flux $F_{\nu,env}$ and width $\sigma_{env}$. Though this is a relatively simple parameterization of what might otherwise be a more complex structure \citep[e.g.][]{Ulrich1976}, the main intention is to provide a reasonable approximation of the large scale data so the ``disk components" of the model need not try to fit such emission and can focus on the disk.

Here, we report 6 models with varying numbers of components and parameters that are representative of the full range of models that we considered:
\begin{enumerate}
    \item Base rectangle model, $\psi = \{x_0, y_0, x_w, y_{w,0}, p.a., \\ \gamma_{in} = \gamma_{out} = 0, F_{\nu}, \sigma_{env}, F_{\nu,env}, A = 0, \\ \gamma_x = \gamma_y = 4\}$.
    \item Broken power-law rectangle model, $\psi = \{x_0, y_0, \\ x_w, y_{w,0}, p.a., \gamma_{in}, \gamma_{out}, x_b, \Delta, F_{\nu}, \sigma_{env}, F_{\nu,env}, \\ A = 0, \gamma_x = \gamma_y = 4, \delta=0.1dx\}$.
    \item Asymmetric broken power-law rectangle model, $\psi = \{x_0, y_0, x_w, y_{w,0}, p.a., \gamma_{in}, \gamma_{out,+}, \gamma_{out,-}, x_b, \Delta, \\ F_{\nu}, \sigma_{env}, F_{\nu,env}, A = 0, \gamma_x = \gamma_y = 4, \delta=0.1dx\}$.
    \item Flared Asymmetric broken power-law rectangle model, $\psi = \{x_0, y_0, x_w, y_{w,0}, p.a., \gamma_{in}, \\ \gamma_{out,+}, \gamma_{out,-}, x_b, \Delta, F_{\nu}, \sigma_{env}, F_{\nu,env}, \gamma_x,\gamma_y, A, \\ \delta=0.1dx\}$.
    \item Gapped Flared Asymmetric broken power-law rectangle model, $\psi = \{x_0, y_0, x_w, y_{w,0}, \\ p.a., \gamma_{in}, \gamma_{out,+}, \gamma_{out,-}, x_b, \Delta, F_{\nu}, \sigma_{env}, F_{\nu,env}, \gamma_x, \\ \gamma_y, A, \delta, x_{gap}, w_{gap}, \Delta_{gap}\}$.
    \item Symmetric Gapped Flared Asymmetric broken power-law rectangle model, $\psi = \{x_0, \\ y_0, x_w, y_{w,0}, p.a., \gamma_{in}, \gamma_{out,+}, \gamma_{out,-}, x_b, \Delta, F_{\nu}, \sigma_{env}, \\ F_{\nu,env}, \gamma_x, \gamma_y, A, \delta, x_{gap}, w_{gap}, \Delta_{gap}\}$.
\end{enumerate}
In the list above, we report only broken power-law models (e.g. Equation \ref{equation:broken_powerlaw}; or in the case of Model 1, Equation \ref{equation:rectangle} with no power-law) because these consistently fit the observations better than a single power-law model, with typical Bayes Factors of $\sim6 - 45$ in favor of the models with the broken power-law intensity profile. We do, however, note that the choice does not impact any of our conclusions. We also note that for simplicity, throughout much of the remainder of the text we will refer to these models by their number in the above list rather than the full name describing the model, though we will also mention the pertinent features of the model along with the number as well.

\begin{deluxetable*}{cc|ccc}
\tablecaption{Best-fit Analytic Model Parameters}
\tablenum{2}
\tabletypesize{\scriptsize}
\label{table:analytic_best_fits}
\tablehead{\colhead{Parameter} & \colhead{Unit} & \colhead{7 mm} & \colhead{1.3 cm} & \colhead{2 cm}}
\startdata
$x_0$ & $''$ & $-0.28406^{+0.00077}_{-0.00085}$ & $-0.2790^{+0.0010}_{-0.0011}$ & $-0.2962^{+0.0029}_{-0.0089}$ \\[2pt]
$y_0$ & $''$ & $-0.1696^{+0.0033}_{-0.0035}$ & $-0.1696^{+0.0030}_{-0.0030}$ & $-0.1505^{+0.0030}_{-0.0031}$ \\[2pt]
$x_w$ & $''$ & $0.1556^{+0.0080}_{-0.0131}$ & $0.1310^{+0.0104}_{-0.0099}$ & $0.128^{+0.020}_{-0.018}$ \\[2pt]
$y_w$ & $''$ & $0.0256^{+0.0024}_{-0.0032}$ & $0.0249^{+0.0045}_{-0.0040}$ & $0.0307^{+0.0064}_{-0.0132}$ \\[2pt]
$p.a.$ & $^{\circ}$ & $182.22^{+0.90}_{-0.90}$ & $182.6^{+1.7}_{-1.1}$ & $193.8^{+4.6}_{-5.8}$ \\[2pt]
$F_{\nu}$ & mJy & $3.55^{+0.12}_{-0.13}$ & $0.821^{+0.043}_{-0.044}$ & $0.466^{+0.029}_{-0.030}$ \\[2pt]
$\gamma_x$ & \nodata & $4.21^{+1.03}_{-0.83}$ & \nodata & \nodata \\[2pt]
$\gamma_y$ & \nodata & $3.35^{+1.25}_{-0.81}$ & \nodata & \nodata \\[2pt]
$A$ & \nodata & $0.96^{+0.28}_{-0.21}$ & \nodata & \nodata \\[2pt]
$\gamma_{in}$ & \nodata & $0.435^{+0.774}_{-0.095}$ & $1.00^{+0.77}_{-0.26}$ & $1.908^{+0.083}_{-0.402}$ \\[2pt]
$\gamma_{out,+}$ & \nodata & $-0.13^{+0.20}_{-0.77}$ & $-0.35^{+0.27}_{-0.54}$ & $-0.84^{+0.59}_{-0.14}$ \\[2pt]
$\gamma_{out,-}$ & \nodata & $0.242^{+0.083}_{-0.765}$ & $0.14^{+0.37}_{-0.41}$ & $1.69^{+0.22}_{-0.90}$ \\[2pt]
$x_b$ & $''$ & $0.0151^{+0.0088}_{-0.0085}$ & $0.019^{+0.017}_{-0.013}$ & $0.0063^{+0.0106}_{-0.0012}$ \\[2pt]
$\Delta$ & \nodata & $8.4^{+70.0}_{-8.0}$ & $0.25^{+1.29}_{-0.24}$ & $2.1^{+7.8}_{-2.1}$ \\[2pt]
$\sigma_{env}$ & $''$ & $0.230^{+0.159}_{-0.067}$ & $0.143^{+0.029}_{-0.014}$ & $0.177^{+0.036}_{-0.026}$ \\[2pt]
$F_{\nu,env}$ & mJy & $0.52^{+0.39}_{-0.21}$ & $0.279^{+0.050}_{-0.055}$ & $0.286^{+0.041}_{-0.041}$ \\[2pt]
\hline
Model No. & Description & \multicolumn{3}{c}{log(Bayes Factor) = $\log{Z_{M_i}} - \log{Z_{M_{Ref.}}}$}\\[2pt]
\hline
1 & Rectangle & $-113.75\pm{0.60}$ & $-252.79\pm{0.53}$ & $-43.35\pm{0.51}$ \\[2pt]
2 & Broken Power-law Rectangle & $-90.86\pm{0.58}$ & $-76.91\pm{0.55}$ & $-29.71\pm{0.52}$ \\[2pt]
3 & Asymmetric Broken Power-law Rectangle & $-55.72\pm{0.60}$ & Ref. & Ref. \\[2pt]
4 & Flared Asymmetric Broken & Ref. & $1.51\pm{0.57}$ & $39.47\pm{0.55}$ \\
  &  Power-law Rectangle  & \multicolumn{3}{c}{  } \\ [2pt] 
5 & Gapped Flared Asymmetric Broken & $3.06\pm{0.60}$ & $2.03\pm{0.57}$ & $42.44\pm{0.55}$ \\
  &  Power-law Rectangle  & \multicolumn{3}{c}{  } \\ [2pt] 
6 & Symmetric Gapped Flared Asymmetric & $1.86\pm{0.62}$ & $1.29\pm{0.57}$ & $41.53\pm{0.56}$\\
  &  Broken Power-law Rectangle  & \multicolumn{3}{c}{  } \\ [2pt] 
\enddata
\end{deluxetable*}

We provide a summary of all of the analytic model parameters, including a short description of what they represent, in Table \ref{table:analytic_priors}. We note that a number of parameters could span orders of magnitude in value, and we fit the base-10 logarithm of the parameter rather than its linear value. For most parameters, we assume a simple uniform prior that limits the value to a reasonable range, but do put further restrictions on some. In particular, we require that the break in the power-law slope occurs ``inside" the rectangle, i.e. $x_b < x_w$, and that the gap falls between the break and the edge, when present. We also require that $\gamma_{in} > 0$ to represent the steep increase in brightness at the center of the disk, but only require that $\gamma_{out} > -1$ to allow for a brightness profile that is flat or even increasing with radius over some portion of the disk. We also require that for the asymmetric models $\gamma_+ > \gamma_-$ so that the posterior is not bi-modal with the brighter half of the disk occurring at either positive or negative $x$ values. When considering size scales in the model, in particular $x_w$, $y_w$, and $x_b$, we use 0\farcs005 as a lower limit on those sizes as features on such scales should be well below the resolution of our observations and therefore difficult to distinguish. For $w_{gap}$ we use a smaller limit of 0\farcs001 as the Gaussian gap is actually wider than this by a factor of a few. We include the priors for each parameter in Table \ref{table:analytic_priors}.

We fit these models to the data directly in the two-dimensional visibility plane, where the errors are best calibrated. To do so, we generate models in the image plane with $1024^2$ pixels that are smaller than the pixels of the observed images in Figure \ref{fig:data} by a factor of 4 (i.e. $dx = 0.25\,dx_{image}$). We then use the \texttt{galario} package \citep{Tazzari2018} to Fourier transform the model image into the visibility plane and sample at the baselines of the observations. 

\begin{figure*}[t]
    \centering
    \includegraphics[width=7in]{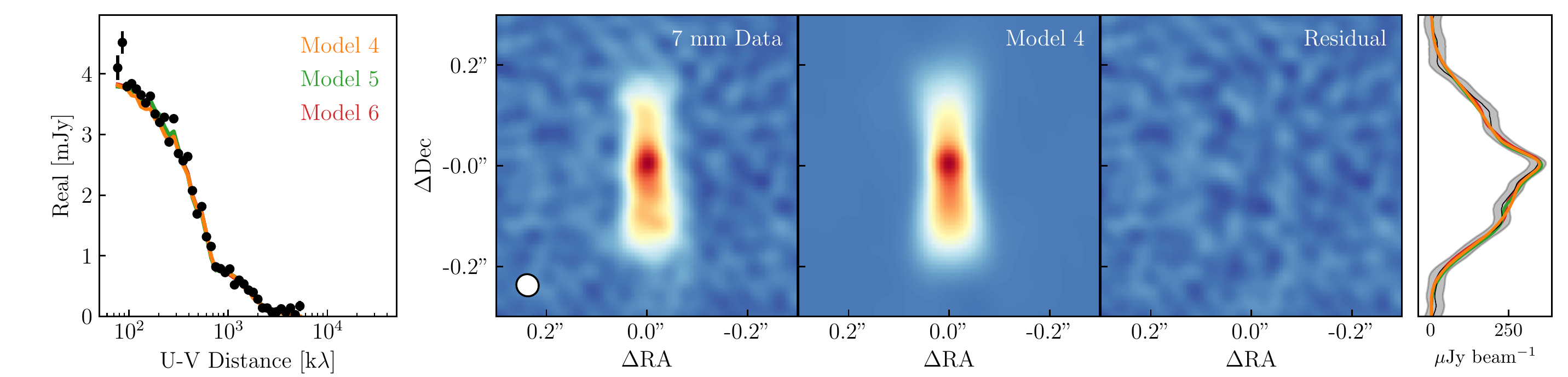}
    \includegraphics[width=7in]{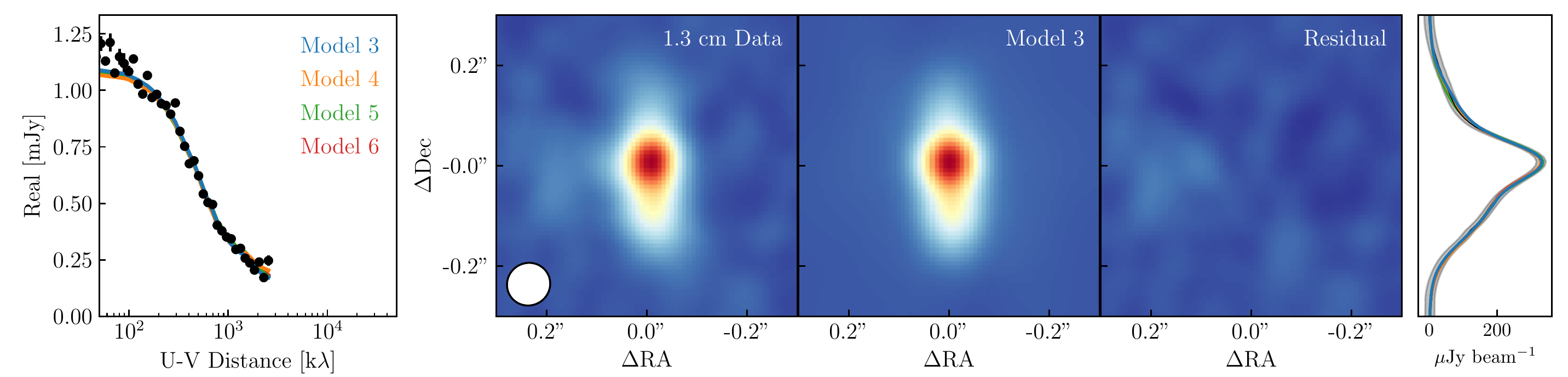}
    \includegraphics[width=7in]{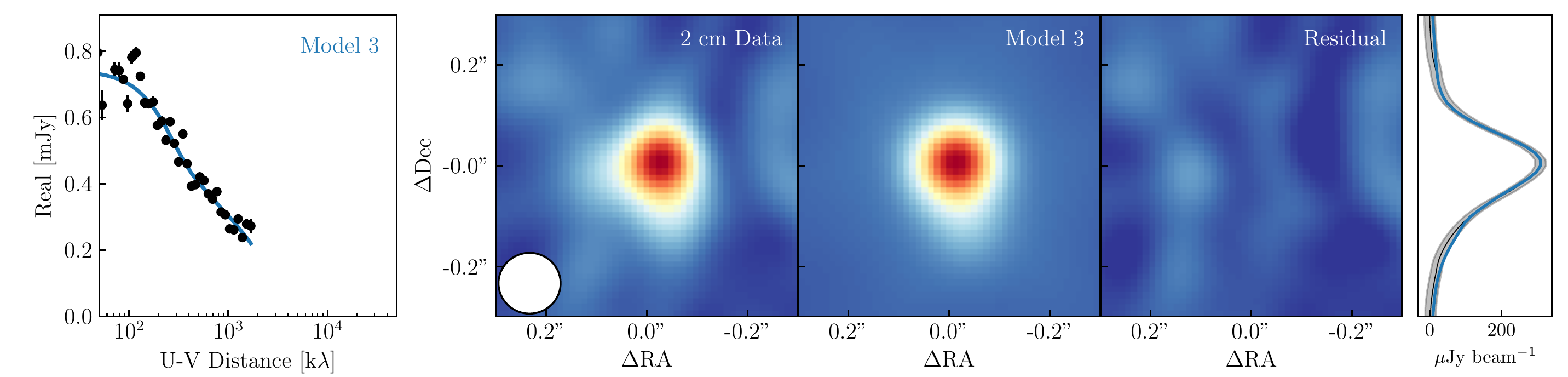}
    \caption{A comparison of our observations of L1527 IRS with our reference model at each wavelength using the maximum likelihood parameters from the respective fit, with the 7 mm data shown in the top row, the 1.3 cm data in the center row, and the 2 cm data on the bottom. The left column shows the one-dimensional azimuthally averaged visibilities compared with the reference model curve. Though the visibilities are shown averaged radially for ease of viewing, all fits were done to the full two dimensional data. In the middle three columns we show the images of our data, the reference model, and the residuals. The model and residual images and visibilities were generated by Fourier transforming a model image, sampling at the same baselines as the data in the $uv$-plane using \texttt{GALARIO}, subtracting these synthetic visibilities from the data in the case of the residuals, and re-imaging with a CLEAN implementation built into the \texttt{pdspy} package. Finally, in the rightmost column we show an intensity profile from a one-dimensional slice through the center of both the data and model images. We also show curves for other models with equivalent Bayesian evidence to our reference model in the leftmost and rightmost panels.}
    \label{fig:best_fit_models}
\end{figure*}

\subsection{Analytic Modeling Results}

\begin{figure*}
    \centering
    \includegraphics[width=7in]{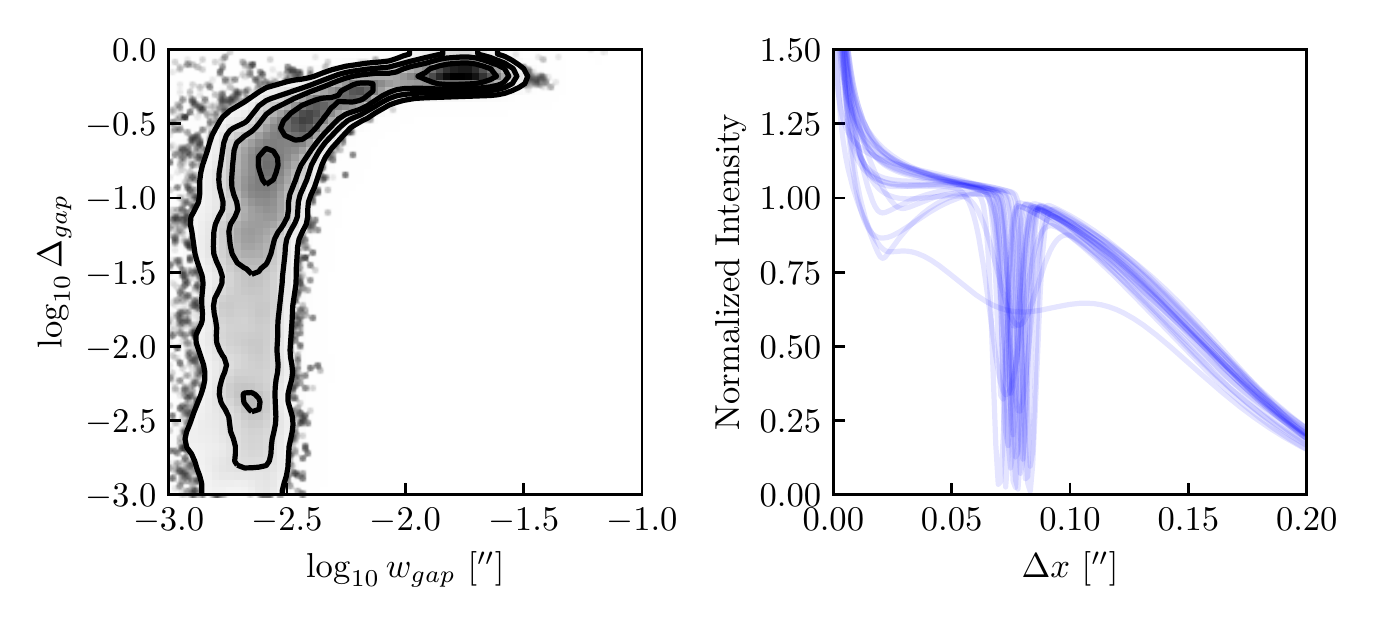}
    \caption{({\it left}) The two-dimensional, marginalized posterior for the $\log w_{gap}$ and $\log \Delta_{gap}$ parameters of our best-fit single-gapped model (Model 7) for the 7 mm data. ({\it right}) 20 realizations of intensity profiles for the L1527 IRS disk at 7 mm drawn from the posterior of the fit, and zoomed in on the gap to see the structure better. We note that the intensity profile is normalized at the location of the gap, before the gap is added, for easier comparison of the gap shape. The solution is quite degenerate, with both narrow and deep or shallow and wide gaps allowed. In the latter case, the feature may be more of a ``shoulder", with the emission dropping and flattening out rather than falling to a true local minimum, instead of an actual ``gap" in the emission.}
    \label{fig:gap_profile}
\end{figure*}

The results of our model fitting are presented in Table \ref{table:analytic_best_fits}. Though we hesitate to define a ``best" model, as our analysis does not always provide a singular best model, for the purposes of presenting a reasonable amount of information we select a ``reference" model from among our model fits. To select our reference model, we choose the simplest model for which the Bayes Factor compared with the previous model indicates strong evidence for the new model (Bayes Factor $>2.5$ following the Jeffrey's scale) with $3\sigma$ significance. We list the parameters for that representative dataset in Table \ref{table:analytic_best_fits}, calculated as the peak of the marginalized posterior for each individual parameter with the uncertainties representing the range around this peak that contains 68\% of the posterior samples. We also show an image of that model compared with the observations in Figure \ref{fig:best_fit_models} using the maximum-likelihood parameters from the fit for that model. We also list the Bayes Factors calculated relative to this representative model in Table \ref{table:analytic_best_fits}.

We find that at 7 mm, Model 4, which includes a broken power-law profile with an asymmetry and flaring, provides the last significant increase in Bayesian Evidence. Models that include an asymmetry (Models 3+) provide a substantial improvement in the quality of fit as compared with models that do not have an asymmetry in the disk brightness (Models 1/2). This improvement demonstrates that the asymmetry that could be made out visually is indeed a statistically significant feature in the data. Including the parameters that control flaring of the disk (Model 4) also provides a strong increase in the Bayesian Evidence as compared with models that do not include flaring (Models 1 -- 3), indicating that the flaring that can be made out in Figure \ref{fig:data} is also real.

We further find that adding either a single gap or a symmetric gap (Models 5/6) produces Bayes Factors that are consistent with zero indicating that we cannot determine whether one of Models 4, 5 or 6 is a better representation of the data than the others. In other words, our observations are perfectly consistent with a gap-free intensity profile, but the presence of gaps in the intensity profile is also allowed. We can, however, use the posterior distributions inferred from fits of models with gaps to characterize the properties of gaps that would be allowed by our observations. In Figure \ref{fig:gap_profile}, we show the posterior distribution for the width and depth of the gap ($w_{gap}$ and $\Delta_{gap}$), marginalized over all other parameters, for the best fit single-gap model (Model 5). To better explore, visually, the range of possible solutions, we also randomly sample 20 models from the posterior distribution and plot the intensity profile for those models along the major axis in Figure \ref{fig:gap_profile}. We find that solutions with a range of gap widths could be consistent with our observations. Wide gaps are required to be quite shallow, otherwise they would have been detectable by our observations. In fact, for the most shallow ``gaps", the feature is barely even present and may be more of a shoulder that drops and flattens out rather than descending to a local minimum, similar to what has been seen in other high resolution imaging of older disks \citep[e.g.][]{Huang2018a}. Narrow gaps, however, particularly those with widths $\sim0\farcs01$, below the resolution of our observations, could potentially be quite deep.

At 1.3 cm and 2 cm we find that Model 3, which includes an asymmetry but no further additional model components, provides the last significant improvement to the quality of fit to both sets of observations. Adding additional components or parameters does not result in a statistically significant improvement in the Bayesian evidence. We note that, strictly speaking, adding flaring to the disk (Models 4 -- 6) improves the quality of the model fit to the 2 cm data substantially. On inspecting those models, however, we found that the flaring of the disk was actually being used to fit the East-West jet feature that can be seen in the 2 cm image. We therefore discard these models but also note that adding the gap feature within them did not increase the Bayesian evidence either, so there does not appear to be evidence for the presence of a gap in the 2 cm observations. This is somewhat unsurprising, however, given the lower resolution and sensitivity to dust emission of these data. As such, we can confidently say that the disk is indeed asymmetric at all three wavelengths of our imaging.

To summarize, we find that disk-like emission is detected at all three wavelengths, and furthermore, models for which the disk is asymmetric provide statistically significant increases in the evidence for those models. This demonstrates, in a statistically rigorous way, that the disk of L1527 IRS is indeed asymmetric with a brighter southern side, as can be seen by-eye in the images in Figure \ref{fig:data}. We find no conclusive evidence, however, that gaps are present in the data. That said, we cannot rule such structures out, but can provide constraints on the sorts of features that might still be consistent with our observations.

\section{Discussion}
\label{section:discussion}

One of the main findings of this work is that the L1527 IRS disk has an asymmetric brightness profile. Interestingly, a similar North-South asymmetry has previously been seen in emission from various molecular lines that have a strong contribution from the disk; for example, $^{13}$CO 2 -- 1 \citep{VanTHoff2018}, C$^{18}$O 2 -- 1 \citep{Aso2017,VanTHoff2018}, C$^{17}$O 2 -- 1 \citep{vantHoff2020TemperatureWarm}, CS 5 -- 4, H$_2$CO 5$_{1,5} - 4_{1,4}$ \citep{Sakai2014}, and c-C$_3$H$_2$ $9_{1,8} - 8_{2,7}$ \citep{Sakai2014a}. Transitions with a strong envelope component display a more symmetric intensity profile, even when a different line for the same molecule traces the disk; for example, CCH 3 -- 2 \citep{Sakai2014}, c-C$_3$H$_2$ $5_{2,3} - 4_{3,2}$ \citep{Sakai2014a}, and CN 2 -- 1 \citep{Tychoniec2021WhichSources}. 

The edge-on nature of the disk makes it difficult to disentangle the underlying physical nature of this feature, however. One would typically assume that thermal dust emission at 7 mm is optically thin, and therefore traces dust surface density. If that was the case then this asymmetry would seem to indicate some enhancement of dust surface density on the southern side of the disk. Such a density enhancement could be related to a number of physical mechanisms that have been seen in more face-on images of protoplanetary disks as well as other protostellar disks, such as vortices or one-armed spirals \citep[e.g.][]{VanDerMarel2013,vanderMarel2016VORTICESDISK,Dong2018TheDisk,Boehler2018TheCavity,Cazzoletti2018EvidenceDisk,Sheehan2020}, that produce pressure bumps and thereby cause dust grains to pile up \citep[e.g.][]{Barge1995DidVortices,Birnstiel2013LopsidedDisks,Meheut2012Dust-trappingDisks}.

That said, with the disk edge-on it is not entirely clear that the disk should be optically thin, even at such long wavelengths, as we may be looking through the entire column of the disk. We do note, though, that the disk asymmetry becomes more pronounced at 1.3 cm than it is at 7 mm, with the brightness a factor of $\sim1.93\times$ brighter $0\farcs1$ to the southern side than to the northern side at  1.3 cm but only a factor of $\sim1.13\times$ brighter at 7 mm. As dust is more optically thin at longer wavelengths, this would be consistent with a scenario where the disk is at least partially optically thin at longer wavelengths and we therefore see deeper into the density enhancement there. If the disk is optically thick, the asymmetry could still be related to a dust density enhancement, though it would need to be far enough out to be in the optically thin region.

To further estimate the optical depth of the 7 mm emission, we compare the brightness temperature of the observations with the estimated temperature profile. We use two separate estimated temperature profiles to do so. First we consider the temperature profile expected for a $\sim2$ L$_{\odot}$ protostar ($T = (L_* / 4\pi\sigma R^2)^{0.25}$). We find that the brightness temperature of the disk around $\sim0\farcs1 = 14$ au of $\sim60 - 75$ K falls below the expected temperature of $\sim125$ K, indicating an optical depth of $\sim0.6 - 0.9$. On the other hand, if we consider the temperature profile from from \citet[][$T = 33 \, (R \, / \, 38.5 \, \mathrm{au})^{-0.35}$ K]{VanTHoff2018}, which is based on measurements of the temperature of L1527 IRS's disk using CO isotopologues, we find an expected temperature of $\sim50$ K, suggesting an optical depth $\sim1$. We note that the temperature profile from \citet{VanTHoff2018} is based on measurements at larger disk radii, where heating from the envelope \citep[e.g.][]{Agurto-Gangas2019RevealingPer-emb-50} is important. This profile may, however, be too shallow when extrapolated inwards to smaller radii where direct heating from the protostar becomes increasingly important. As such, this latter value is likely an upper limit on the optical depth. Collectively, it seems likely that the disk is perhaps partially optically thin at 7 mm, and increasingly optically thin at longer wavelengths, consistent with the appearance of the asymmetry across images at these wavelengths.

Another interesting result is that the well-resolved 7 mm emission appears to trace the north-south oriented disk well, indicating that it is dominated by dust thermal emission rather than free-free emission. If this is true, it would imply that a significant amount of relatively large grains already exists in this deeply embedded Class 0 disk since mm/cm-sized grains (sometimes referred to as ``astrophysical pebbles") are generally thought to be optimal for producing dust emission at VLA Bands.

We estimate the amount of dust present in the disk using the standard assumption of optically thin dust emission such that the dust mass can be calculated from the millimeter flux, \citep[e.g.][]{HILDEBRAND1983}:
\begin{equation}
    M_{d} = \frac{F_{\nu} \, D^2}{B_{\nu}(T) \, \kappa_{\nu}}.
\end{equation}
We use a distance of 140 pc \citep[e.g.][]{Torres2007VLBATaurus,Zucker2019} and a temperature of 51 K typical of protostellar disks \citep[e.g.][]{Tobin2020} based on a suite of radiative transfer models that find that average protostellar disk temperatures follow $T = 43 \, \mathrm{K} (L / L_{\odot})^{0.25}$ and a $\sim2$ $L_{\odot}$ protostar \citep{Kristensen2012WaterProtostars}. For the flux of the disk, we use the disk flux estimated from the rectangle model (Model 1), of 3.3 mJy. The rectangle model without the power-law component to match the central peak that is likely dominated by free-free emission should provide the best estimate of the disk-only flux from our modeling. The dust opacity is the largest source of uncertainty, as it depends significantly on the dust grain size distribution. We assume that
\begin{equation}
    \kappa_{\nu} = 2.3 \left(\frac{\nu}{\mathrm{230 \, GHz}}\right)^{\beta} \mathrm{cm^2 \, g^{-1}},
\end{equation}
where we adopt a 230 GHz opacity of 2.3 cm$^2$ g$^{-1}$ following \citet{Andrews2013}, and where small, micron-sized grains typical of the interstellar medium have $\beta\approx1.5-2$ while grains grown to sizes similar to the observed wavelength have $\beta\approx0$ \citep[e.g.][]{Hartmann2008MassesDisks}. We find that, depending on the value of $\beta$, L1527 IRS has between 15 and 411 M$_{\oplus}$ for $\beta=0$ and $\beta=2$, respectively. This is lower than the dust mass found by \citet{Nakatani2020}, of $\sim866$ M$_{\oplus}$, likely due to the difference in 7 mm dust opacity; for $\beta=0$ our dust opacity is 0.08 cm$^{2}$ g$^{-1}$ while theirs is 0.02 cm$^{2}$ g$^{-1}$, though the different ways that we treat the temperatures may also play a role. Despite the differences in masses, our recovered fluxes are in good agreement, with \citet{Nakatani2020} finding a total flux of 3.7 mJy to our 3.6 mJy when including the central peak. The mass found by \citet{Tobin2013} from more careful modeling of 870 $\mu$m and 3.4 mm observations, of $\sim25$ M$_{\oplus}$ is in good agreement with the lower end of our range.

It is interesting to note that the 7 mm emission is vertically extended, with a best fit width of 14 au assuming a distance of $\sim140$ pc \citep[e.g.][]{Torres2007VLBATaurus,Zucker2019}, implying that the grains responsible for its emission have yet to settle to the disk midplane. We would naively expect that the large mm/cm-sized dust grains that are primarily probed by our 7 mm observations should settle to the midplane on a timescale relatively short compared with the age of L1527 IRS. We estimate that the scale height of the gas, calculated as $h = c_s / \Omega$ at $\sim25$ au with $c_s$ using both temperature profiles described previously, is $\sim2.3 - 3.5$ au, depending on temperature profile, or a width of $\sim5 - 7$ au, consistent with the dust extending vertically up to a few scale heights in the disk. The timescale for settling is given by \citep[see, e.g., the][review article on``Physical Processes in Protoplanetary Disks", Section 7.2]{Armitage2015PhysicalDisks}:
$$
t_{\rm settle}= {\rho\over \rho_m} {v_{\rm th}\over s} {1\over \Omega_{\rm K}^2} = {\rho\over \rho_m} {v_{\rm th}\over s} {R^3\over G M_*} 
$$
$$
= 5.7\times 10^3 \left({\rho\over 10^{-12} {\rm g\ cm}^{-3} }\right) \left({ 3~{\rm g\ cm}^{-3}\over \rho_{\rm m}  }\right) \left({ 0.1\ {\rm cm} \over s  }\right) 
$$
$$
\times \left({ T\over 50 {\rm K} }\right)^{1/2} \left({ R\over {25\ {\rm au}}  }\right)^3 \left({ {0.45\ {\rm M}_\odot}\over M_* } \right)\ {\rm years},
$$
where $\rho$ is the local gas density, $\rho_{\rm m}$ the dust material density, $v_{\rm th}$ the gas thermal speed, $\Omega_{\rm K}$ disk rotation angular frequency, $R$ the local radius, and $M_*$ the central stellar mass. 

The most uncertain quantity is the gas density at $R\sim 25$~au near the outer edge of the 7~mm disk. It is constrained by the Toomre parameter
$$
Q={c_s \Omega_{\rm K}\over \pi G \Sigma}={M_*\over 2\pi \rho R^3}
$$
$$
=2.7 \left({10^{-12} {\rm g\ cm}^{-3} \over \rho }\right) \left({ M_*\over {0.45 {\rm M}_\odot} }\right) \left({ 25\ {\rm au}\over R }\right)^3.
$$
It would be difficult for the gas density to go well above the fiducial value of $10^{-12}$~g~cm$^{-3}$, which corresponds to a Toomre Q value that is already close to unity. As such, any dust settling time we calculate above is likely an upper limit on the true timescale for dust to settle.

For the fiducial values of the gas density and other quantities, the dust settling time is about $6,000$~years for mm-sized grains (and 10 times shorter for cm-sized grains), which is significantly shorter than the time scale for the Class 0 ($\sim 1.6\times 10^5$~years) and I ($\sim 5.4\times 10^5$~years) stages of star formation \citep[e.g.][]{Evans2009,Dunham2015}. One would therefore expect such large grains to have settled to the midplane unless they are continuously stirred up by some kind of ``turbulent" flows in the disk meridional plane. The fact that the large grains emitting at 7~mm do not appear to be settled may indicate the presence of a significant level of disk turbulence, which would be consistent with active accretion that is needed to transport the fast envelope infall through the disk to the central protostar, through mechanisms such as the magneto-rotational instability \citep[e.g.][]{Balbus1991}. 

Finally, we also note that \citet{Bae2018Planet-drivenImplications} found that the number of spiral arms driven in a disk by a planet is determined, in part, by the disk's aspect ratio ($h/r$, where $h$ is the scale height and $r$ is the radius within the disk). They found that higher $h/r$ typically led to only a single spiral in the outer disk. If the relatively large vertical extent of the large grains in this disk, as evidenced by the East-West extent, is indicative of similarly high values of $h/r$ for the gas, this would then be consistent with the presence of a single spiral arm in the disk driving the asymmetry. 

It is, of course, quite speculative to assume a planetary origin for the asymmetry, and indeed there are other plausible mechanisms by which such asymmetries might be formed. 
Single spiral arms are to be expected for massive disks with large aspect ratios where self-gravity is dominant, leading to local gravitational instabilities within the disk \citep[e.g.][]{Kratter2016GravitationalDisks}. Infall from envelope to disk may also drive spiral arms in the disk \citep[e.g.][]{Tomida2017}, or alternatively could incite Rossby wave instabilities that form vortices within the disk \citep[e.g.][]{Bae2015AREINFALL}. Asymmetric ``streamers" of infalling material that have recently been found towards some protostars \citep[e.g.][]{Alves2020AFormation,Pineda2020ALength,Thieme2022Accretion3-MMS} could also preferentially deposit material asymmetrically into the disk. Interestingly, the ALMA Band 4 observations of L1527 IRS shown in \citet{Nakatani2020} show what appears to be a slight asymmetry on the {\it north}western side of the disk. Based on its location high in the disk and towards the outskirts, this feature could also be associated with infall, if real. Regardless of the origin, however, such over-densities of material could potentially serve as sites with conditions favorable for planet formation.

\section{Conclusions}
\label{section:conclusion}

In this work, we have presented new, high sensitivity VLA observations at 7 mm, 1.3 cm, and 2 cm of the Class 0 protostar L1527 IRS. Our observations show an edge-on protostellar disk visible at all three wavelengths, along with a central point-like feature that increases in brightness relative to the disk at longer wavelengths. This central point source is likely a combination of both dust emission as well as free-free emission associated with a jet, with the contribution from the jet increasing at longer wavelengths, and indeed an East-West protrusion can be seen perpendicular to the disk at 1.3 cm and 2 cm. With our new, order of magnitude higher sensitivity observations, we do not find evidence of the clumps reported by \citet{Nakatani2020}, instead finding that the disk is gap-free to the limit of our sensitivity and resolution. We do find, however, that the disk is asymmetric at all three wavelengths, with the southern half of the disk appearing brighter than the northern half.

To confirm these features visible in the image-plane, we conduct careful modeling of the observations in the $uv$-plane. We find that models that include an asymmetry in the disk provide statistically significant improvements to the Bayesian evidence in favor of those models compared with models that do not include an asymmetry at all three wavelengths, indicating that the asymmetry is indeed a real feature of the observations. We also find that at 7 mm, models that include flaring of the disk  provide statistically significant increases in evidence in favor of those models. Models that include a gap feature {have equivalent Bayesian evidence to gap-less models. As such, we cannot rule out that such features might be present in the intensity profile, though our modeling does provide constraints on the properties of such putative features.}

The origin of the asymmetry is unclear, and particularly difficult to interpret due to the edge-on appearance of the disk, but could be associated with spiral arms or vortices in the disk, or other features that might produce an asymmetric density enhancement. The large vertical extent of the disk is consistent with simulations of both planet-disk interactions and also gravitationally unstable disks. Such a spiral could, in turn, produce an asymmetry like the one seen here. Infall from envelope to disk could viably produce such a feature, as well. The large vertical extent inferred for the large dust grains in the disk also suggests a significant level of disk turbulence, consistent with active accretion through the disk at early times.

Regardless of the origin of the asymmetry, its presence provides an interesting look at the conditions in a particularly young disk around a single protostar, where such observations are sorely lacking. Further observations like these will be critical for understanding the onset of planet formation in the youngest disks.

\software{CASA \citep{Mcmullin2007}, dynesty \citep{Speagle2019}, matplotlib \citep{Hunter2007}, corner.py \citep{Foreman-Mackey2016}}, galario \citep{Tazzari2018}

\begin{acknowledgements}
We would like to thank the anonymous referee whose feedback helped to focus and clarify the manuscript.
P.D.S is supported by a National Science Foundation Astronomy \& Astrophysics Postdoctoral Fellowship under Award No. 2001830.
J.J.T. acknowledges funding from NSF grant AST-1814762.
Z.Y.L. is supported in part by NASA 80NSSC18K1095 and NSF AST-1910106. 
M.L.R.H. acknowledges support from the Michigan Society of Fellows.
J.K.J. and S.G. acknowledge support from the Independent Research Fund Denmark (grant No. 0135-00123B).
L.W.L. acknowledges support from NSF AST-2108794.
N.O. acknowledges support from the Ministry of Science and Technology (MOST) of Taiwan (MOST 109-2112-M-001-051 and MOST 110-2112- M-001-031).
S.T. is supported by JSPS KAKENHI grant Nos. 21H04495 and 21H00048. 
J.P.W. acknowledges support from NSF grant AST-2107841.
C.W.L. is supported by Basic Science Research Program through the National Research Foundation of Korea (NRF) funded by the Ministry of Education, Science and Technology (NRF-2019R1A2C1010851).
K.T. is supported by JSPS KAKENHI Grant Numbers JP16H05998 and JP21H04487.
 
The National Radio Astronomy Observatory is a facility of the National Science Foundation operated under cooperative agreement by Associated Universities, Inc.
\end{acknowledgements}

\bibliographystyle{aasjournal}
\bibliography{references.bib}{}

\appendix

\section{Comparison with Nakatani et al. (2020)}
\label{section:nakatani_comparison}

\subsection{Archival VLA Observations}

To compare the results of our new observations with the results presented in \citet{Nakatani2020}, we re-reduce the archival VLA observations 
from that work, focusing on the Q and K-band data taken in A (Program 11A-188) and B configuration (program 13A-401).
Extensive details of the observations are
presented in \citet{Nakatani2020}, but the A configuration
data were taken in 2011 with 2 GHz of bandwidth, and the B-configuration 
data were taken in 2013 with 8 GHz of bandwidth. The pointing center was set at $\alpha$(J2000)$\,=\,04^{\rm h}39^{\rm m}53.6^{\rm s}$ $\delta$(J2000)$\,=\,26^{\circ}03{\arcmin}06.0{\arcsec}$ for both sets of observations. 
A key difference to point out is that different complex gain calibrators 
were used for Q-band (J0438+3004) and K-band (J0431+2037) in the A configuration 
observations, which, as we discuss later, may have enhanced the appearance of the ``clumps" that were previously reported.

The data were reduced using the scripted version of the VLA pipeline in CASA 4.2.2, following the same procedures for data editing as the new data presented here.
The data could not be processed with the most current pipeline as they were obtained during the commissioning phase of the VLA in 2011. At Q-band we performed additional flagging of the data following \citet{Nakatani2020} in order to match their data as best as possible. This included flagging the data from 2011 which was observed at elevations $<35^{\circ}$ as well as baselines taken on 2011 August 06 with projected baseline lengths $>1000$ k$\lambda$. To correct for proper motion between the 2011 and 2013 epochs noted by \citet{Nakatani2020}, we created images of each epoch individually using the $tclean$ routine within CASA with multi-frequency synthesis mode, a robust parameter of 2.0, and a 1000 k$\lambda$ taper to smooth the data to approximately the same resolution. We then fit each epoch separately with a two-dimensional Gaussian in the image plane and used the centroid from those fits to align the epochs at the same location.

We produce a final image of the data using the $tclean$ routine, employing multi-frequency synthesis mode and a robust parameter of -1.0. The resulting Q-band image has a beam size of $0\farcs079 \times 0\farcs056$ with a position angle of $-85.0^{\circ}$ and an RMS of 60 $\mu$Jy beam$^{-1}$. We note that \citet{Nakatani2020} used a robust parameter of -2.0, but we opt for a less aggressive value as we found that decreasing beyond this increased the noise of the image too steeply. Our image does, however, qualitatively resemble the image from their work, particularly with regards to the ``clumpy" structures that were reported. To more quantitatively compare our re-imaging with \citet{Nakatani2020}, we convert our image from units of Jy beam$^{-1}$ to brightness temperature, and find a peak near $\sim100$ K and that the brightness temperatures of the northern and southern shoulders of the disk fall near 60-70 K, in agreement with what was previously found. We further note that the image presented in \citet{Nakatani2020} used a robust parameter of -2 and had an rms of 110 $\mu$Jy beam$^{-1}$, while our reproduction of their work uses a robust parameter of -1 and has a correspondingly lower rms of 60 $\mu$Jy beam$^{-1}$. As such, the image presented in \citet{Nakatani2020} is $\sim10\times$ less sensitive than our new observations while our reproduction is only $\sim7\times$ less sensitive. Our K-band image has a beam size of $0\farcs099 \times 0\farcs092$ with a position angle of $-53.2^{\circ}$ and an RMS of 28 $\mu$Jy beam$^{-1}$. Both images are shown in Figure \ref{fig:old_image}. As was done for our new observations, we also scale the weights on the visibility data by a factor of 0.125 to match the RMS of a naturally weighted image of the data.

\subsection{Comparison of Archival Data with Our Observations}
\label{section:comparison}

\citet{Nakatani2020} previously reported the detection of three clumps in the disk of L1527 IRS with Q-band imaging. The clumps were labelled ``N", ``C", and ``S", are located on the northern side of the disk, near the center of the disk, and on the southern side of the disk, respectively and can be seen in this imaging.. Our Q-band image presented in Figure \ref{fig:data} has $\sim10\times$ lower rms than the image presented in their work, and although we cannot rule out substructures in the disk as well, the restrictions on such features are, at first glance, qualitatively quite different from the large clumps that were previously reported. 

\begin{figure*}
    \centering
    \includegraphics[width=7in]{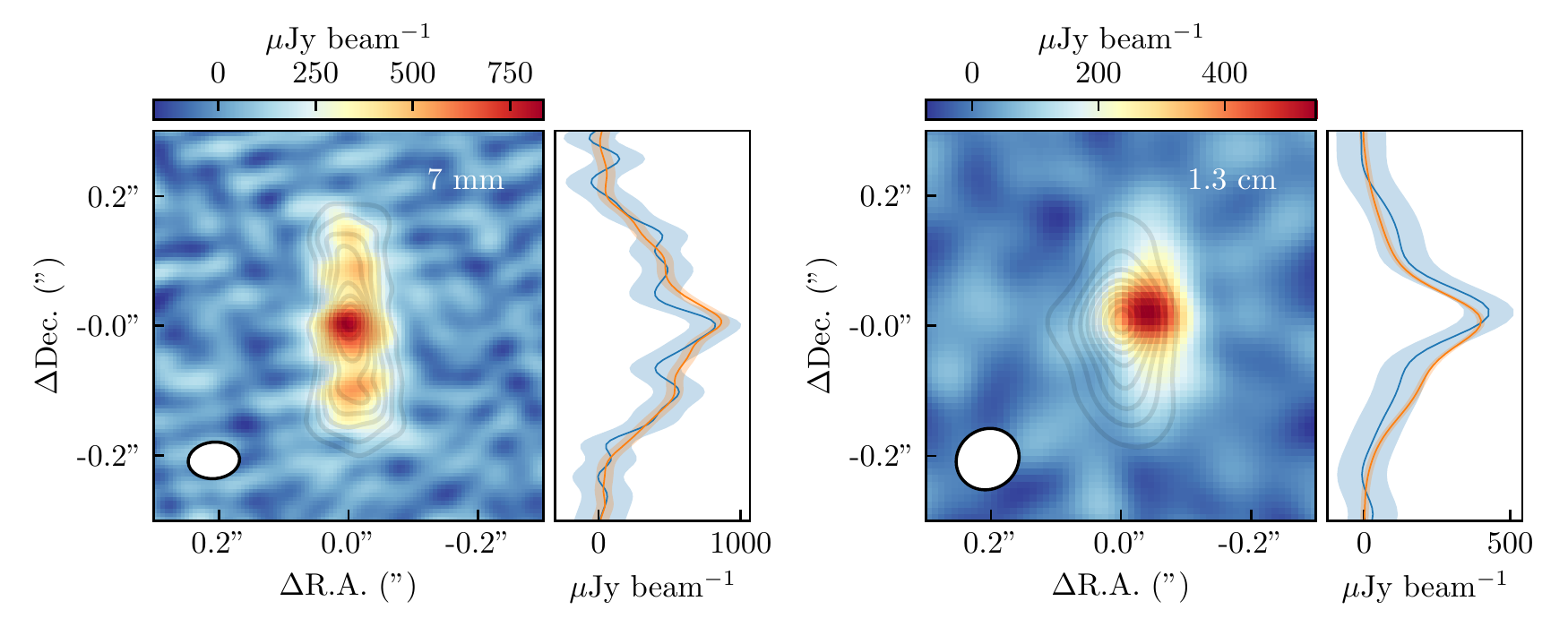}
    \caption{({\it left}) Robust = -1.0 weighted image of L1527 IRS at 7 mm using the observational data presented in \citet{Nakatani2020}. Though not identical to the image presented in \citet{Nakatani2020}, the ``clumpy" features that they find in their image are reproduced qualitatively here. ({\it right}) Robust = 0.5 weighted image of L1527 IRS at 1.3 cm using the observational data presented in \citet{Nakatani2020}, but centered at the same position as the 7 mm to demonstrate the offset between Q-band and K-band. In both images we show the intensity profile from a one-dimensional pixel slice along the North-South direction through the center of the disk to the right of the image in blue, along with a shaded region representing the $3\sigma$ confidence interval. The same one-dimensional slice for our new observations, scaled by a factor of the ratio between old and new beam sizes to account for difference in beam area, is shown in orange for a direct comparison with our newer data. We also show contours at intervals of 5$\sigma$ ({\it 7 mm}) and 10$\sigma$ ({\it 1.3 cm}) for our new observations with significant transparency so as to not obscure the underlying image.}
    \label{fig:old_image}
\end{figure*}

The C clump found by \citet{Nakatani2020} likely corresponds to the central peak in our own imaging. That said, this feature is not likely a clump or substructure so much as the inner region of the disk with some contribution from free-free emission associated with a jet. As bright central emission remains present out to longer wavelengths, and indeed an east-west jet can be seen at 2 cm, it is likely that this central component is at least in part free-free emission.

Though our modeling cannot rule out that gap features are present on either the north or south side of the disk (models with gaps (Models 5/6) have comparable Bayesian Evidence to gapless models (Model 4) for our 7 mm observations) the restrictions that are placed on such features are qualitatively quite different from the large clumps proposed by \citet{Nakatani2020}. As the sensitivity of our new image is a factor of $10\times$ higher than those presented previously, and the previously reported clumps were $\sim2\sigma$ peaks above the background of the disk, we should have detected such features with high significance in our data. Such features would likely be relatively wide and deep, i.e. down and to the right in Figure \ref{fig:gap_profile}, which we can confidently rule out. 

Instead, we believe that the substructures previously reported are the result of noise peaks or troughs in a noisy image combined with poorer coverage of the $uv$-plane. We note that the flux difference between the gap and the clump is $\sim2\sigma$ when measured at the highest intensity value within the gap. The total L1527 IRS disk has an area of $\sim15$ beams, so assuming Gaussian noise statistics we would expect $0.025 \times 15 = \sim0.4$ noise peaks (or troughs) with a $>2\sigma$ significance within the disk. We would also expect numerous $\sim1\sigma$ peaks or troughs that could work together to create the appearance of clumps.

The clumps seen by \citet{Nakatani2020} were likely further emphasized by the subtraction of the compact central free-free emission seen in their K-band observations when imaged with robust=-2 weighting. Doing so may have enhanced the appearance of the clumps in two ways: first there is a 
systematic spatial offset of $0\farcs05$ between the Q-band emission and K-band emission in the 
\citet{Nakatani2020} data. As the astrometry of interferometric images is tied to the position of the gain calibrator, which was different for the Q- and K-band observations presented in \citet{Nakatani2020}, systematic offsets in the positions of said calibrators could lead to systematic astrometric offsets between two images of the same source but made with different calibrators. As there is no such offset between our new Q- and K-band images that employed the same phase calibrator between them (and also different from either calibrator used in \citet[][]{Nakatani2020}), the offset is likely due to the different phase calibrators used between the Q- and K-band observations from \citet{Nakatani2020}. The subtraction of the K-band source performed by \citet{Nakatani2020} was done without correcting for this offset and over-emphasized clump C. Moreover, both our new K-band image (see Figure \ref{fig:data}) and also the previous data (see Figure \ref{fig:old_image}) show extended emission from the disk. Thus, at the same time, the subtraction of the K-band image also likely
subtracted both dust and free-free emission and not only free-free emission.

\begin{figure*}
    \centering
    \includegraphics[width=7in]{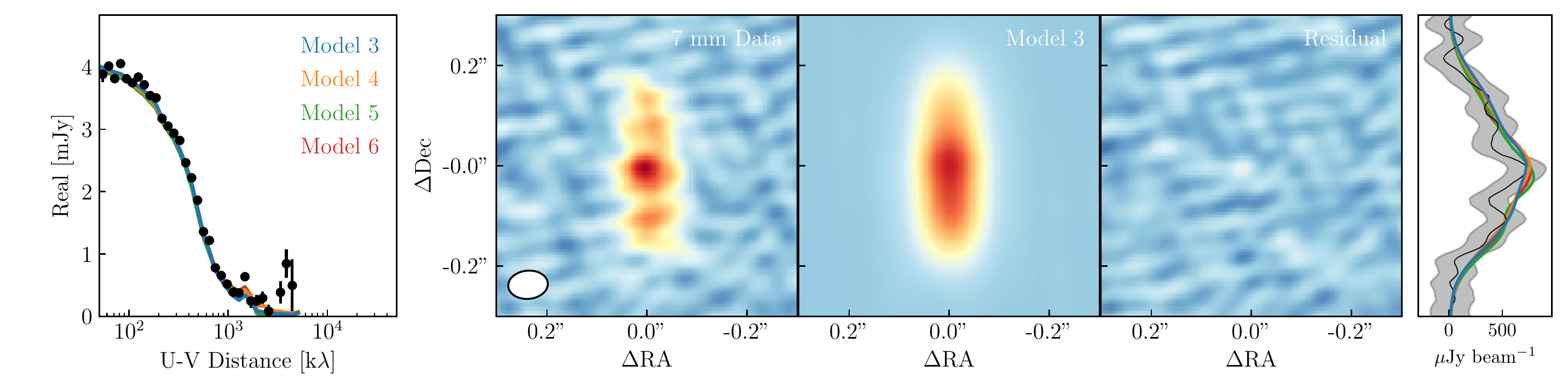}
    \caption{A comparison of the observations of L1527 IRS from \citet{Nakatani2020} at 7 mm with the reference model using the maximum likelihood parameters from the respective fit. The left column shows the one-dimensional azimuthally averaged visibilities compared with the reference model curve. Though the visibilities are shown averaged radially for ease of viewing, all fits were done to the full two dimensional data from \citet{Nakatani2020}. In the middle three columns we show the images we made from those data, the reference model, and the residuals. The model and residual images and visibilities were generated by Fourier transforming a model image, sampling at the same baselines as the data in the $uv$-plane using \texttt{GALARIO}, subtracting these synthetic visibilities from the data in the case of the residuals, and re-imaging with a CLEAN implementation built into the \texttt{pdspy} package. Finally, in the rightmost column we show an intensity profile from a one-dimensional slice through the center of both the data and model images. We also show curves for other models with equivalent Bayesian evidence to our reference model in the leftmost and rightmost panels.}
    \label{fig:old_best_fit_model}
\end{figure*}

Finally, to test for the presence of gaps in the intensity profile quantitatively, we repeat our modeling analysis for the previous observations. Following the same conventions regarding the reference model for this fit, we list the results of this modeling in Table \ref{table:old_analytic_best_fits} and we show a comparison of the reference model with the observations in Figure \ref{fig:old_best_fit_model}. We find that Model 3, which includes an asymmetry but no further components, provides the last significant strong increase in Bayesian evidence. This suggests that with these older data it would have been possible to confidently identify the disk as having a North-South asymmetry. While the addition of further components does increase the Bayesian evidence in some cases, these increases are typically insignificant when compared with the previous model. The most significant increase as compared with the reference model is Model 5, which includes flaring and a gap, with Bayes factor of $1.56 \pm 0.55$ that does not provide strong evidence in favor of that model, particularly not when compared with other more extensive models (e.g. Model 4). We cannot, however, rule out models with either flaring, a gap, or both, either. Indeed, the posteriors from gapped-disk models find constraints on the gap features that might be present that are similar to our own observations, though somewhat less constraining due to the lower sensitivity of the observations.

\begin{deluxetable}{cc|c}
\tablecaption{Best-fit Analytic Model Parameters for Archival Data}
\tablenum{3}
\tabletypesize{\scriptsize}
\label{table:old_analytic_best_fits}
\tablehead{\colhead{Parameter} & \colhead{Unit} & \colhead{7 mm}}
\startdata
$x_0$ & $''$ & $3.7384^{+0.0013}_{-0.0014}$ \\[2pt]
$y_0$ & $''$ & $3.6307^{+0.0117}_{-0.0087}$ \\[2pt]
$x_w$ & $''$ & $0.1593^{+0.0083}_{-0.0111}$ \\[2pt]
$y_w$ & $''$ & $0.0388^{+0.0039}_{-0.0048}$ \\[2pt]
$p.a.$ & $^{\circ}$ & $181.32^{+1.05}_{-0.99}$ \\[2pt]
$F_{\nu}$ & mJy & $3.57^{+0.11}_{-0.13}$ \\[2pt]
$\gamma_{in}$ & \nodata & $0.38^{+0.65}_{-0.13}$ \\[2pt]
$\gamma_{out,+}$ & \nodata & $-0.19^{+0.18}_{-0.73}$ \\[2pt]
$\gamma_{out,-}$ & \nodata & $0.12^{+0.14}_{-0.64}$ \\[2pt]
$x_b$ & $''$ & $0.0162^{+0.0099}_{-0.0100}$ \\[2pt]
$\Delta$ & \nodata & $9.8^{+64.3}_{-9.8}$ \\[2pt]
$\sigma_{env}$ & $''$ & $0.290^{+0.119}_{-0.080}$ \\[2pt]
$F_{\nu,env}$ & mJy & $0.56^{+0.15}_{-0.16}$ \\[2pt]
\hline
Model No. & Description & log(Bayes Factor) \\[2pt]
\hline
1 & Rectangle & $-28.30\pm{0.52}$ \\[2pt]
2 & Broken Power-law Rectangle & $-12.58\pm{0.54}$ \\[2pt]
3 & Asymmetric Broken Power-law Rectangle & Ref. \\[2pt]
4 & Flared Asymmetric Broken & $0.65\pm{0.55}$ \\
  &  Power-law Rectangle  &  \\ [2pt] 
5 & Gapped Flared Asymmetric Broken & $1.56\pm{0.55}$ \\
  &  Power-law Rectangle  &  \\ [2pt] 
6 & Symmetric Gapped Flared Asymmetric & $0.91\pm{0.55}$\\
  &  Broken Power-law Rectangle  &  \\ [2pt] 
\enddata
\end{deluxetable}

\end{document}